# The Tian Pseudo Atom Method

Hui LI*, Meng HE* and Ze ZHANG

Beijing University of Technology, Beijing 100124, People's Republic of China

CAS Key Laboratory of Nanosystem and Hierarchical Fabrication, National Center for Nanoscience and Technology, Beijing 100190, People's Republic of China

Zhejiang University, Hangzhou 310014, People's Republic of China

Correspondence e-mail: huilicn@yahoo.com, mhe@nanoctr.cn

**Abstract**

In this work, the authors gives a new method for phase determination, the pseudo atom method (PAM). In this new method, the figure of merit function, $R_{Tian}$, replaces the normal $R_{CF}$ function in charge flipping algorithm. The key difference between $R_{CF}$ function and $R_{Tian}$ function is the observed structure factor was replaced by the pseudo structure factor. The test results show that $R_{Tian}$ is more powerful and robust than $R_{CF}$ to estimate the correct result especially with low resolution data. Therefore, the pseudo atom method could overcome the charge flipping method's defeat to some extent. In theory, the pseudo atom method could deal with quite low resolution data but this needs a further test.

## 1. Introduction

The phase problem in crystallography could be converted into a global optimization problem [Harrison, 1990; Prince, 1993]. In our previous work [Li, *et al.*, 2015], we gave two definition evaluation functions to estimate the correct solution from the whole possible solutions. They are Tian1 and Tian2 functions. To high resolution data, Tian1 function is a quite good figure of merit (FOM) to estimate the correctness of the solution. At the same work, Tian1 function gives the reason why charge flipping algorithm (CF) [Oszlányi & Sütó, 2004] can make success. But, when the data resolution decreased, Tian1 function will be defeat. Of course, CF algorithm will also be defeat with low resolution data. To overcome this problem, we gave Tian2 function. Unfortunately, we found two shortcomings for Tian2 function in the further test. One is that it needs maximum entropy calculation in Tian2 function. This needs a great computation resource. The other is Tian2 function is just correct for the equal-atom structure. These two shortcomings will prevent Tian2 function from being used in structure determination.

In this work, we gave a new method to deal with the phase problem for the low resolution data. We call this method Tian pseudo atom method (TPAM), or pseudo atom method (PAM) for short. In this paper, we will introduce iteration Fourier algorithms, figure of merit (R value) of these algorithms and R value defeat under low resolution data in part 2. The pseudo structure factor was introduced to deal with R value defeat in this part either. In part 3, to calculate pseudo structure factor, pseudo atom scatter factor was introduced. In this part, we also introduce how to use Wilson statistics theory to calculate pseudo structure factor from the observed structure factor, atom scatter factor and pseudo atom scatter factor. In part 4, the flow chart of the pseudo atom method was given. Two calculation examples were given in part 5 which show the valid of the PAM. The last part is the discussion and conclusion of the PAM.

## 2. Iteration Fourier algorithms and pseudo structure factor

In 2004, Oszlányi and Sütó introduced charge flipping (CF) algorithm into crystallography field [Oszlányi & Sütó, 2004]. Up to now, this algorithm has become very popular due to its simplicity, high computation efficiency and validity [van Smaalen, *et al.*, 2003; Spek, 2003, Palatinus & Chapuis, 2007; Coelho, 2007; Oszlányi & Sütó, 2008]. In nature, CF method is one kind of iteration Fourier algorithms just like hybrid input and output algorithm (HIO) and error reduced (ER) algorithm [Fienup, 1982]. Unfortunately, these algorithms will be defeat when the data resolution decreased. Figure 1 shows the basic flow chart of CF algorithm. In CF algorithm, $R_{CF}$ value was used as figure of merit (FOM) to estimate the convergence of the iterate calculation. The $R_{CF}$ is defined [Oszlányi & Sütó, 2004] as follows:

$$R_{CF} = \frac{\sum_h \left| |F_{h,obs}| - |F_{h,cal}| \right|}{\sum_h |F_{h,obs}|} \tag{1}$$

where $F_{obs}$ is the observed structure factor. $F_{cal}$ is the calculation structure factor after density modification (DM) (charge flipping in CF method). Due to density modification in every iteration course, $|F_{obs}|$ would never be equal to $|F_{cal}|$. But it proved by practical that with high resolution data, the solution which has lowest $R_{CF}$ value approximately equals to the correct structure. Unfortunately, when the data resolution decreased, this approximation doesn't valid anymore. In our previous work shows that all zero solution (all phases of structure factors are set to zero degree) always having a lower $R_{CF}$ value than the correct solution [Li, *et al.*, 2015]. The range of data resolution which ensures Tian1 valid is about 1.2 Å - 1.8 Å. This range is determined by the element type, Dybe-Waller factor and so on.

Now, let's inspect Eq.1 again. If we could know the correct solution (correct phases) in advance, then we can calculate the correct calculation structure factor $F_{cal}^{correct}$ directly. See Figure 1. When we have correct phase, we could calculate $|F_{cal}^{correct}|$ with one time Fourier transform, density modification and inverse Fourier transform. Using $|F_{cal}^{correct}|$ to replace $|F_{obs}|$ in Eq.1, then Eq.1 will be convert into:

$$R^* = \frac{\sum_h \left| |F_{h,cal}^{correct}| - |F_{h,cal}| \right|}{\sum_h |F_{h,cal}^{correct}|} \tag{2}$$

In theory, the $R^*$ value of the correct solution should be zero and the other incorrect solution will have no-zero value. Therefore, the defeat of CF method under low resolution data will be overcome. For convenience, we call the correct calculation structure factor $F_{cal}^{correct}$ is the pseudo structure factor $F_{PS}$. The reason is that $F_{obs}$ comes from the real structure (real crystal) but $F_{PS}$ from the real structure after density modification which was called pseudo structure. Now, the question is how to obtain the $|F_{PS}|$ when the structure is unknown.

## 3. Pseudo structure factor and pseudo atom scatter factor

The structure factor has relationship with the atom scatter factor as follows:

$$F(H) = \sum_{i}^{N} f_i(H) \times \exp(-2\pi i \overline{H} \bullet \overline{R_i}) \qquad (3)$$

where $F(H)$ is structure factor. $f_i(H)$ is the *i*th atom's scatter factor. H is Miller index. $\overline{R_i}$ is the *i*th atom's coordinate.

According to Eq.3, pseudo structure factor should have:

$$F_{PS}(H) = \sum_{i}^{N} f_{i,PS}(H) \times \exp(-2\pi i \overline{H} \bullet \overline{R_i}) \qquad (4)$$

where $F_{PS}(H)$ is the pseudo structure factor. $f_{i,PS}(H)$ is the pseudo atom scatter factor. In other word, if we want to obtain the pseudo structure factor, we must acquire the pseudo atom scatter factor at first.

According to Eq.3, if there was only one atom in the unit cell and set this atom into the original point of the lattice, the structure factor of this single atom crystal will be equal to this atom's scatter factor. It was deduced as follows:

$$\because F(H) = \sum_{i}^{N} f_i(H) \times \exp(-2\pi i \overline{H} \bullet \overline{R_i})$$

Let $N = 1$ and $\overline{R_i} = (0,0,0)$

$$\therefore F(H) = f \qquad (5)$$

At the same way, when we make density modification after we got charge density of the single atom crystal, we could obtain pseudo charge density of this crystal, $\rho^{*}_{crystal}$. Doing inverse Fourier transform (IFT), we will obtain the pseudo structure factor of this single atom crystal, $F_{PS}$. Base on Eq.5, this pseudo structure factor equals to pseudo atom scatter factor, $F_{PS} = f_{PS}$ . Figure 2 shows the calculation course of the pseudo atom scatter factor. Figure 3 gives the atom scatter factor and pseudo atom scatter factor of the carbon in $C_{252}H_{326}O_{19}$ crystal. The data resolution is set to 2.5 Å.

After obtained the pseudo atom scatter factor, but we could not calculate pseudo structure factor directly using Eq.4. Because we could not get atoms' coordinates, $\overline{R_i}$, before crystal structure determined. Therefore, we use Wilson statistics theory to deal with it. According to Wilson's theory [Wilson, 1949]:

$$\left\langle |F_{obs}|^2 \right\rangle = \sum_i^N f_i^2 \tag{6}$$

$\langle\ \rangle$ means expected value. Of course, the pseudo structure factor and pseudo atom scatter also obey this relationship:

$$\left\langle |F_{PS}|^2 \right\rangle = \sum_i^N f_{i,PS}^2 \tag{7}$$

From Eq.6 and Eq.7, it obtains:

$$|F_{PS}| = |F_{obs}| \times \sqrt{\frac{\sum_i^N f_{i,PS}^2}{\sum_i^N f_i^2}} \tag{8}$$

Eq.8 shows that if we had observed structure factor, atom scatter factor and pseudo atom scatter factor, the pseudo structure factor could be calculated directly. After obtained pseudo structure factor, using it replace the observed structure factor in R value calculation (Eq.2). This is so-called Tian pseudo atom method (TPAM) or pseudo atom method (PAM).

## 4. The flowchart of the pseudo atom method

The main calculation flowchart of pseudo atom method as follows:

Step 1: Collect diffraction data, and determine the lattice parameters.

Step 2: Determine the element type and calculate every element's atom scatter factor.

Step 3: Calculate every element's pseudo atom scatter factor according to the Figure 2.

Step 4: According to Eq.8, calculate pseudo structure factor from the observed structure factor, atom scatter factor and pseudo atom scatter factor.

Step 5: Set different phase values $\varphi_H$ to $|F_{obs}(H)|$ . Then calculate charge density, ρ and modification charge density, $\rho^*$ of the structure. At last, calculate $|F_{cal}(H)|$. The whole course is as follows:

$$|F_{obs}| + \varphi_H \xrightarrow{FT} \rho \xrightarrow{DM} \rho^* \xrightarrow{IFT} |F_{cal}|$$

Step 6: Calculate figure of merit (FOM). We call this FOM is $R_{Tian}$. It is defined:

$$R_{Tian} = \sum_h \frac{\left| |F_{h,PS}| - |F_{h,cal}| \right|}{w_h} \tag{9}$$

and the R is defined:

$$R = \sum_h \frac{\left\| F_{h,obs} \right| - \left| F_{h,cal} \right\|}{w_h} \quad (10)$$

where $w_h$ is the weight factor. In this work, all $w_h$ were set to 1.

## 5. Examples

GaN

The lattice parameters of GaN are a=b=3.18 Å, c=5.189 Å, α = β = 90.0 °, γ = 120.0 °. There are two Ga atoms and two N atoms in one unit cell. The data resolution was set to 1.0 Å. The calculation results were given in Table 1. Column 2 is the theory observed structure factor. Column 3 is pseudo structure factor which calculated with Eq.8. Column 4 is theory pseudo structure factor which calculated with Eq.4. These two pseudo structure factors are quite consistent which shows that Eq.8 is correct. The last three columns are calculated structure factors in three different cases. The first is the correct solution. The second is all zeros solution and the last is the lowest $R_{CF}$ value (or the highest Tian1 value) solution. The lowest $R_{CF}$ result is not correct because it just gives two Ga atoms positions but none N atoms positions. So it was called wrong result. In terms of R value, the wrong result has the lowest value (R=75.1) which is lower than the correct result's (R=75.9). But in terms of $R_{Tian}$, the correct result has the lowest value ($R_{Tian}$=25.6) and the wrong result has a higher value ($R_{Tian}$=29.3).

$C_{252}H_{326}O_{19}$ [Czugler, et al, 2003]

This structure is quite hard to be determined because it has none heavy atom and the unit cell is quite large. Our pervious work shows CF method will defeat to determine this structure when the data resolution decreased to about 1.5 Å. In this work, we set the data resolution to 2.5 Å. The lattice parameters are $a$=16.909 Å, $b$=18.772 Å, $c$=21.346 Å, $\alpha$=111.46° , $\beta$= 103.38° , $\gamma$=107.74 ° . Figure 3 gives the carbon's atom scatter factor and its pseudo scatter factor. Table 2 gives four different cases $R_{Tian}$ and R values. The first result is the correct structure result. The second is the all zeros result. The third one is the lowest $R_{CF}$ value result. The last one is the random phase result. In the last case, the phase of the structure factors was set from 0° to 360° randomly and obey Friedel's law. From these results, it shows that in terms of R value, both incorrect result (wrong result, R=11127.59) and all zero result (R=14376.63) have lower R values than the correct result (R=14858.40). But in terms of $R_{Tian}$ , the correct result has the lowest value ($R_{Tian}$=7097.39). This means using PAM and $R_{Tian}$ could estimate the correct result when R or Tian1 defeat.

These two examples show that pseudo atom method could overcome the CF method's defeat problem quite efficiently.

## 6. Discussion and conclusion

The core idea of the pseudo atom method is shown in Figure 4. The crystal was made up of atoms. In CF/ER/HIO algorithms, the modification charge density of the crystal ( $\rho^{*}_{cryst}$ ) was obtained by modified the whole real crystal's charge density ( $\rho^{real}_{cryst}$ ). On the other hand, we could modify the real atom's charge density ( $\rho^{real}_{atom}$ ) to get the pseudo atom's charge density ( $\rho^{PS}_{atom}$ ) at first. Then, use these pseudo atoms to make up whole pseudo crystal. Therefore, whole pseudo crystal's charge density ($\rho_{cryst}^{PS}$) was obtained. To set up pseudo atom method, we hypotheses these two charge densities are approximately equal, $\rho_{cryst}^{*} \cong \rho_{cryst}^{PS}$. In other word, the valid of PAM was determined by the valid of $\rho_{cryst}^{*} \cong \rho_{cryst}^{PS}$.

In $R^*$ function (Eq.2), if we could get $|F_{cal}^{correct}|$ in advance, $R^*$ could be used to estimate the correct result under any data's resolution condition. On the same way, the valid of the $R_{Tian}$ also has nothing to do with the data's resolution unless $\rho_{cryst}^{*} \cong \rho_{cryst}^{PS}$ was invalid. Therefore, in theory PAM could be used to deal with very low resolution data, but this needs further test.

Compared with Tian2 function, PAM does not need maximum entropy calculation. Therefore, PAM has quite high computation efficiency.

Furthermore, the tests results we have done show that $H_{hy}$ maybe a promising FOM which is defined (see Table 2):

$$H_{hy} = \frac{R_{Tian}}{R} \tag{11}$$

At last, iteration Fourier algorithms will be invalid in PAM. Therefore, it needs develop a high efficient global optimization algorithm in the next step.

The work was financially supported by the National Natural Science Foundation of China (No. 11574060 and 11474281). The authors are grateful to Prof. Yuhui Dong (Institute of High Energy Physics, China) for his suggestion and discussion.

**Figures**

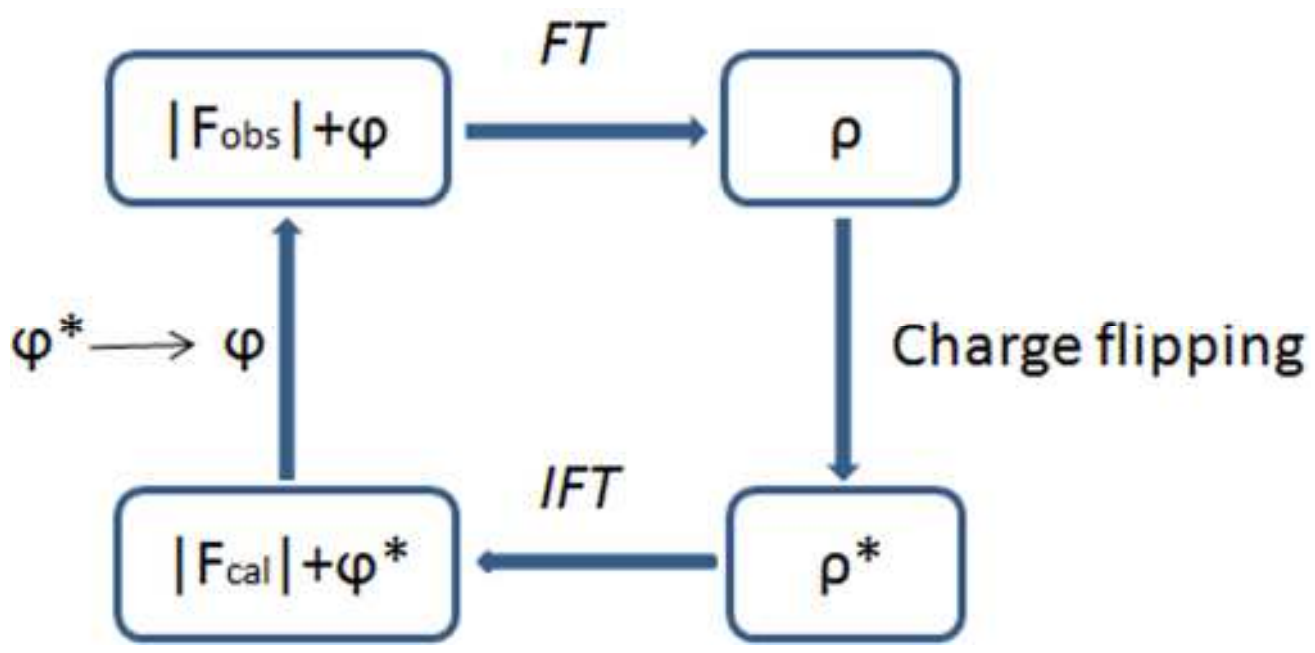

Figure 1. The flowchart of the charge flipping algorithm. FT means Fourier transform. IFT means inverse Fourier transform. In HIO or ER algorithms, charge flipping course was replaced with density modification (DM).

$$F(H) = f(H) \xrightarrow{FT} \rho_{cryst} \xrightarrow{DM} \rho^{*}_{cryst} = \rho^{PS}_{atom} \xrightarrow{IFT} F_{PS}(H) = f_{PS}(H)$$

Figure 2. The flowchart of the pseudo atom scatter factor calculation

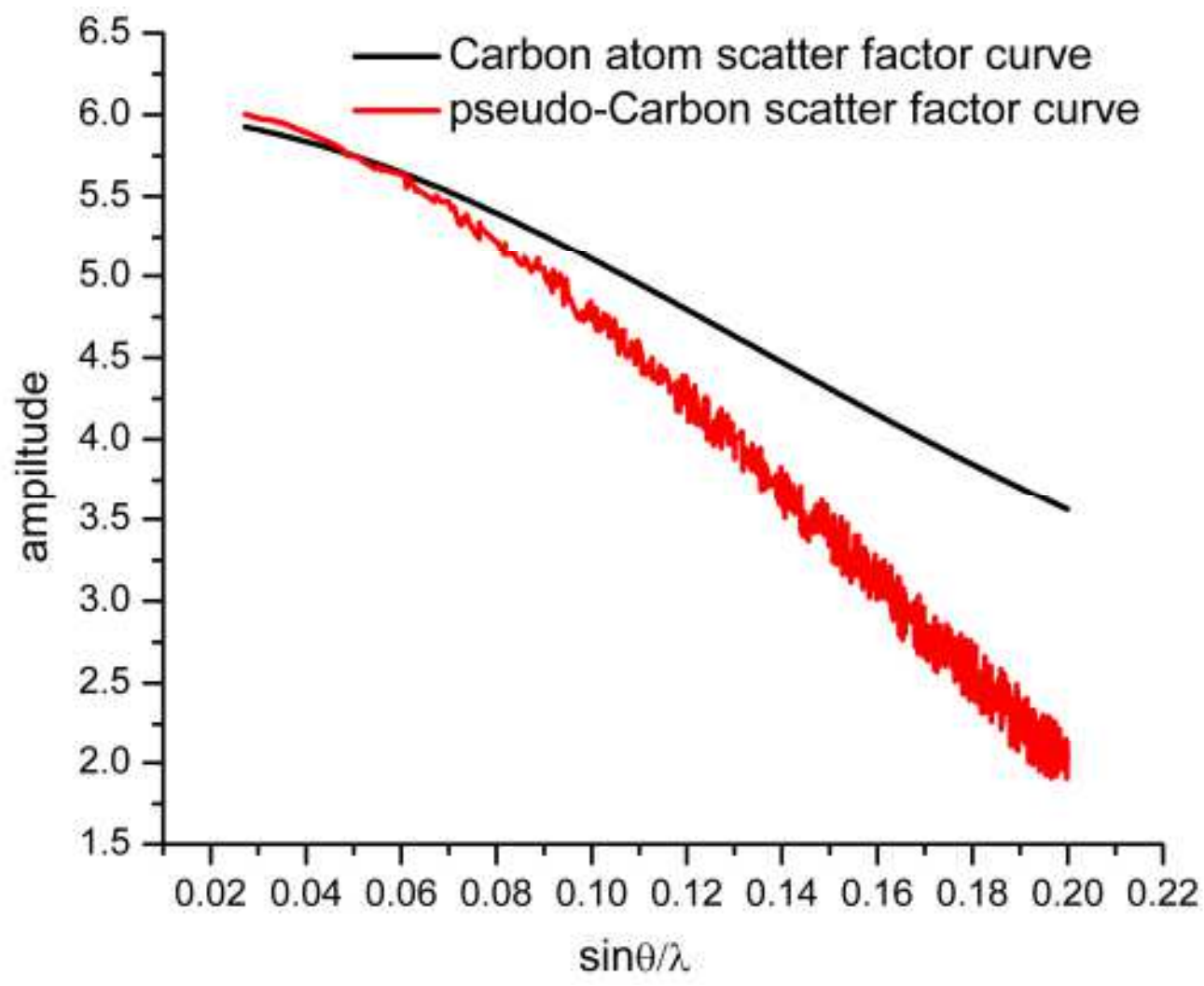

Figure 3. The atom scatter factor and its pseudo scatter factor curves of the carbon atom in $C_{252}H_{326}O_{19}$. The data resolution is 2.5 Å

$$\rho^{real}_{atom} \xrightarrow{\sum} \rho^{real}_{cryst} \xrightarrow{DM} \rho^{*}_{cryst} \cong \rho^{PS}_{cryst} \xleftarrow{\sum} \rho^{PS}_{atom} \xleftarrow{DM} \rho^{real}_{atom}$$

Figure 4. The core idea of the pseudo atom method.

Table 1

| Miller | $\lvert F_{obs}^{theory} \rvert$ | $\lvert F_{PS} \rvert$ | $\lvert F_{PS}^{theory} \rvert$ | $\lvert F_{cal} \rvert$ | | |
|---|---|---|---|---|---|---|
| | | | | ϕ=correct | ϕ=0 | ϕ=wrong |
| (001) | 0.0 | 0.0 | 0.0 | 6.3668E-6 | 12.620 | 0.042 |
| (100) | 30.465 | 30.562 | 30.52 | 31.724 | 28.945 | 31.107 |
| (002) | 51.219 | 50.632 | 50.632 | 50.489 | 55.44 | 50.855 |
| (101) | 37.745 | 37.554 | 37.564 | 42.886 | 41.495 | 43.956 |
| (102) | 22.987 | 21.907 | 21.892 | 22.156 | 20.611 | 22.281 |
| (003) | 0.0 | 0.0 | 0.0 | 4.105E-6 | 1.265 | 0.040 |
| (110) | 48.178 | 44.092 | 44.782 | 44.461 | 46.386 | 43.707 |
| (111) | 0.023 | 0.021 | 0.0 | 0.02 | 5.592 | 0.142 |
| (103) | 38.478 | 32.970 | 33.374 | 31.934 | 29.924 | 31.477 |
| (200) | 22.047 | 18.698 | 18.991 | 18.096 | 16.054 | 17.775 |
| (112) | 38.848 | 32.162 | 32.162 | 31.122 | 34.078 | 31.365 |
| (201) | 29.987 | 24.703 | 24.579 | 25.027 | 24.337 | 25.555 |
| (004) | 32.465 | 25.134 | 24.464 | 27.654 | 31.671 | 28.537 |
| (202) | 18.055 | 13.632 | 13.652 | 13.939 | 11.884 | 14.053 |
| (104) | 15.224 | 10.664 | 10.371 | 14.191 | 12.428 | 14.566 |
| (113) | 0.024 | 0.017 | 0.0 | 0.0176 | 3.584 | 0.023 |
| (203) | 31.024 | 19.809 | 20.192 | 21.998 | 20.793 | 21.768 |
| (210) | 17.930 | 11.223 | 11.557 | 12.121 | 12.071 | 12.055 |
| (005) | 0.0 | 0.0 | 0.0 | 2.4604E-6 | 0.215 | 0.020 |
| (211) | 24.849 | 14.904 | 14.66 | 18.245 | 17.554 | 18.48 |
| (114) | 27.037 | 14.838 | 14.573 | 17.434 | 19.82 | 17.818 |

$R_{Tian} = \sum \left\| F_{PS} \right| - \left| F_{cal}^{\varphi=correct} \right\| = 25.6$ $\quad R_{Tian} = \sum \left\| F_{PS} \right| - \left| F_{cal}^{\varphi=0} \right\| = 64.7$ $\quad R_{Tian} = \sum \left\| F_{PS} \right| - \left| F_{cal}^{\varphi=wrong} \right\| = 29.3$

$R = \sum \left\| F_{obs}^{theory} \right| - \left| F_{cal}^{\varphi=correct} \right\| = 75.9$ $\quad R = \sum \left\| F_{obs}^{theory} \right| - \left| F_{cal}^{\varphi=0} \right\| = 102.2$ $\quad R = \sum \left\| F_{obs}^{theory} \right| - \left| F_{cal}^{\varphi=wrong} \right\| = 75.1$

Table 2

| **$R_{Tian}$** | | | |
|---|---|---|---|
| Φ (correct) | ϕ (all zero) | ϕ (wrong) | ϕ (random) |
| ***7097.39*** | 9862.23 | 7685.98 | 11437.14 |
| **R** | | | |
| Φ (correct) | ϕ (all zero) | ϕ (wrong) | ϕ (random) |
| 14858.40 | 14376.63 | ***11127.59*** | 19302.53 |
| **$H_{hy} = R_{Tian}/R$** | | | |
| Φ (correct) | ϕ (all zero) | ϕ (wrong) | ϕ (random) |
| ***0.478*** | 0.686 | 0.691 | 0.593 |